\documentclass[aps,pra,twocolumn,showpacs,groupedaddress]{revtex4}  

\usepackage[american]{babel}
\usepackage{graphicx}  
\usepackage{dcolumn}   
\usepackage{bm} 
\usepackage{amssymb}   
\usepackage{amsmath}
\usepackage{color} 

\def\op#1{{\Hat{\mathrm{#1}}}}

\def\bra#1{\ensuremath{\langle{#1}\vert}}
\def\ket#1{\ensuremath{\vert{#1}\rangle}}

\def\commutator#1#2{\mathinner{\mathopen[#1,#2\mathclose]}}

\newcommand{\be}{\begin{equation}}
\newcommand{\ee}{\end{equation}}
\newcommand{\benn}{\begin{equation*}}
\newcommand{\eenn}{\end{equation*}}

\newcommand{\beq}{\begin{eqnarray}}
\newcommand{\eeq}{\end{eqnarray}}

\def\H1{\widehat{H}_1}

\newcommand{\lb}{\left[}
\newcommand{\rb}{\right]}
\newcommand{\lp}{\left(}
\newcommand{\rp}{\right)}

\let\oldsqrt\sqrt
\def\sqrt{\mathpalette\DHLhksqrt}
\def\DHLhksqrt#1#2{%
\setbox0=\hbox{$#1\oldsqrt{#2\,}$}\dimen0=\ht0
\advance\dimen0-0.2\ht0
\setbox2=\hbox{\vrule height\ht0 depth -\dimen0}%
{\box0\lower0.4pt\box2}}

\def\H1{\widehat{H}_1}

\begin{document}

\title{Dynamically generated darkness in the Dicke model}
\author{Michael Tomka, Dionys Baeriswyl and Vladimir Gritsev}
\affiliation{Physics Department, University of
Fribourg, Chemin du Mus\'ee 3, 1700 Fribourg, Switzerland}

\date{October 7, 2013}

\begin{abstract}

We study the dynamics of a driven Dicke model, where the collective
spin is rotated with a constant velocity around a fixed axis.
The time evolution of the mean photon number and of the atomic
inversion is calculated using, on the one hand, a numerical technique
for the quantum dynamics of a small number of two-level atoms, on the
other hand, time-dependent mean-field theory for the limit of a large
number of atoms.
We observe a reduction of the mean photon number as compared to its
equilibrium value.
This dynamically generated darkness is particularly pronounced
slightly above the transition to a superradiant phase.
We attribute the effect to a slowing down of the motion in the
classical limit of a large ensemble and to an interplay of dynamic and
geometric phases in the quantum case.

\end{abstract}

\maketitle

\section{Introduction}

The Dicke model~\cite{Dicke1954},~\cite{TC} describes $N$ two-level
atoms interacting with a single-mode radiation field through the
dipole coupling.
If the atoms are confined to a region of space which is much smaller
than the wavelength of the radiation field, the two-level emitters
behave as a single large spin.
The model is a paradigm for the collective behavior of matter
interacting with light.
In the thermodynamic limit, $N\rightarrow\infty$, there is a
transition to a superradiant phase at a critical coupling
strength~\cite{HL},~\cite{WH},~\cite{CGW}.
The order parameter can be chosen to be the photon density, which
vanishes in the normal phase and is finite in the superradiant phase.
Alternatively one may also use the fraction of excited atoms, which
behaves similarly to the photon density. 
Recently the emphasis was on the quantum phase transition at zero
temperature.
It could be shown that the entanglement between photons and atoms
diverges by approaching the quantum-critical point~\cite{Lambert}. 
Interestingly, for finite $N$ clear signatures of the onset of quantum
chaos are also found for couplings at and not too far above the
quantum-critical point~\cite{Emary2003}.

The existence of this phase transition in a real material was called
seriously into question some time ago.
It was argued that adding the term proportional to $A^2$ to the Dicke
model, where $A$ is the vector potential of the electromagnetic field,
would eliminate the equilibrium transition~\cite{RWZ}.
However, in a more recent study it was pointed out that within a
consistent treatment of interactions between excited atoms a
transition would still occur, although maybe of a different type than
originally thought~\cite{Keeling},~\cite{VD}. 
A different route would be to realize the Dicke model in a
non-equilibrium setting of cavity quantum
electrodynamics~\cite{Dimer}.
An experimental breakthrough was achieved with a high-quality cavity
hosting a Bose-Einstein condensate~\cite{Esslinger}.
This setup is not only considered to be a realization of the Dicke
model, it also produced data showing evidence for a superradiant
phase transition and other collective behaviors~\cite{Esslinger_2}.

The combination of cold atomic gases and cavity quantum
electrodynamics has the advantage that parameter values can be tuned
with high precision and even controlled in real time.
This type of experiments has stimulated studies of non-equilibrium
effects and, especially, their interplay with collective phenomena,
for instance in the vicinity of a quantum phase transition.
A fundamental problem is the evolution of an initial state -- or
density matrix -- in a model without dissipation. 
For the (time-independent) Dicke model an approach to equilibrium has
been found, governed by a competition between classical chaos and
quantum diffusion~\cite{AH}. 
The Dicke model turns into a many-body Landau-Zener model if both
atomic and photonic excitation energies become time-dependent. 
In contrast to the usual Landau-Zener model with a single two-level
system, where an adiabatic evolution - and thus no tunneling - is
easily achieved for slow enough driving, this becomes very difficult
in the problem involving many two-level systems~\cite{AGKP}.
At the critical point an adiabatic evolution appears to be
impossible~\cite{IT}.
A different scenario is obtained if the coupling constant is
time-dependent.
Here a rich new phase diagram including metastable phases and
first-order transitions has been reported~\cite{BERB}.

In several papers dealing with the dynamics of the Dicke model the
rotating wave approximation has been used, where the emission of a
photon is accompanied by the transition of an excited atom to its
ground state and the absorption of a photon is linked to the
excitation of an atom.
This yields an integrable model where the dynamics is not chaotic.
In one study the evolution after an initial quench has been found to
lead either to a constant photon density at long times or to produce
persistent oscillations with several incommensurable
frequencies~\cite{Yuzbashyan}.
Other studies investigated the conditions under which a classical
treatment of the dynamics is valid~\cite{TL},~\cite{Faribault}.

In this paper we consider a specific time-dependent coupling where the
spin operator is rotated with a time-dependent angle.
This ``rotated Dicke Hamiltonian'' can be generated by a unitary
transformation starting from the time-independent Hamiltonian.
Therefore the energy eigenvalues do not depend on time, in contrast to
the study mentioned above~\cite{BERB}.
Nevertheless the evolution is highly nontrivial and leads to
qualitatively new effects at and slightly above the equilibrium
quantum critical point.
We are particularly interested in the thermodynamic limit, where a
time-dependent mean-field theory is expected to yield exact results
for the dynamics.
We develop this theory on the basis of a coherent state representation
of both photon and spin degrees of freedom.
We obtain a classical dynamical system of two oscillators, which are
coupled by a nonlinear time-dependent term.
Alternatively, this classical limit can also be reached in the
framework of the Holstein-Primakoff transformation.
Once the classical dynamical system is established, it is
straightforward to calculate the relevant physical quantities, such as
the (time-dependent) photon density.
For finite $N$ we complement the mean-field calculations by a
numerical procedure where the quantum evolution for a specific initial
state is computed for the rotated Dicke Hamiltonian.
For large $N$, these numerical results approach the mean-field
predictions, as expected.

In our calculations we kept all coupling terms of the Dicke model and
did not use the rotating wave approximation.
At weak coupling one does not find noticeable effects of the terms
neglected by the rotating wave approximation, but at and above the
quantum critical point the differences between integrable and
non-integrable cases show up clearly, for instance through the onset
of chaos in the latter case. 

The following general picture emerges from our calculations.
The rotation by a time-dependent angle leads to an upshift of the
critical coupling strength above which a finite photon density is
produced spontaneously.
If the system is prepared in the ground state of the
(time-independent) Dicke model slightly above its quantum-critical
point, i.e. with a finite photon density, an interesting darkening
effect occurs after the rotation is switched on.
A clear minimum in the photon density is found as a function of the
coupling strength for a fixed driving velocity or, even more
pronounced, as a function of driving velocity for a fixed coupling
strength.
The location of the minimum defines a dynamical critical line in the
coupling versus velocity plane.
The effect can be readily understood as a slowing down of the
nonlinear dynamics in the classical limit.
For a small number of two-level systems, where the evolution has to be
treated quantum mechanically, we find that the geometric phase
associated with the time-dependent state of the system plays a crucial
role in this dynamically generated darkness.

The paper is organized as follows.
Section~\ref{sec:timeindependentdickemodel} gives a short review of
the standard (time-independent) Dicke model.
The time-dependent Dicke model is introduced in Section~\ref{sec:timedependentdickemodel}. 
The rotation by a time-dependent angle can be partly compensated by a
transformation to a co-rotating frame.
The numerical procedure used for treating the quantum dynamics of a
limited number of two-level systems is also explained.
Section~\ref{sec_mfa} presents the time-dependent mean-field theory using
coherent states for both the spin and the photons.
The stationary states in the co-rotating frame yields a modified phase
diagram, as compared to that of the time-independent Dicke model.
Two linear modes describe motions close to equilibrium, one of which
softens by approaching criticality.
However, in the vicinity of the critical point the dynamics is highly
nonlinear.
This becomes clear in Section~\ref{darkness} where the evolution of the
system is studied after the rotation is suddenly switched on.
The phenomenon of dynamically generated darkness, observed close to
criticality, is interpreted as a slowing down of the motion, in
analogy to an effect experienced by a particle moving in the Mexican
hat potential.
Section~\ref{compgeomdyn} shows that in the quantum limit (small number of
two-level systems) the minimum in the photon density disappears if the
geometric phase is removed by hand, which suggests that the darkening
effect results from an interplay of geometric and dynamic phases.
A brief summary is presented in the concluding Section~\ref{sec:conclusion}.
Two appendices give some details for the linearized classical dynamics
and the slowing down in the Mexican hat, respectively.

\section{Time-independent Dicke model}
\label{sec:timeindependentdickemodel}

The Dicke model takes into account a single radiation mode and reduces
the atoms to two-level systems, described by Pauli matrices
$\op{\sigma}_{i}^{\mu}$, $\mu=z,\pm$, $i=1,...,N$.
The parameters of the model are the photon frequency $\omega$, the
level splitting $\omega_0$ (we choose $\hbar=1$) and the coupling
constant $\lambda$.
The Hamiltonian reads 
\be
  \label{eq:dickeham}
  \op{H}_{\mathrm{D}} =
  \omega_{0} \op{J}_{z} +
  \omega \op{a}^{\dagger}\op{a} +
  \frac{\lambda}{\sqrt{N}}
  \lp \op{a}^{\dagger} + \op{a} \rp
  \lp \op{J}_{+} + \op{J}_{-} \rp,
\ee
where $\op{J}_{\mu}=\sum_{i=1}^{N} \op{\sigma}_{i}^{\mu}/2$ are
collective atomic operators satisfying the commutation relations of
the angular momentum, 
\be
  \commutator{\op{J}_{+}}{\op{J}_{-}} = 2 \op{J}_{z}, \quad
  \commutator{\op{J}_{z}}{\op{J}_{\pm}} = \pm \op{J}_{\pm}
\ee
and $\op{a}^\dag$, $\op{a}$ are photon creation and annihilation
operators with bosonic commutation relations. 

The Hilbert space of the two-level systems is spanned by the Dicke
states $\ket{j,m}$, which are eigenstates of ${\op{\bf J}}^{2}$ and
$\op{J}_{z}$ with eigenvalues $j(j+1)$ and $m$, respectively, $j$
being a non-negative integer or a positive half-integer and $m$ being
restricted to $-j \le m \le j$. 
$\op{H}_{\mathrm{D}}$ commutes with $\op{\bf J}^2$ and is therefore
block-diagonal with respect to the Dicke states.
As in Ref.~\cite{Emary2003} we fix $j$ to its maximal value $j=N/2$.
This is a natural choice because the ground state for $\lambda=0$
belongs to this subspace. 
The collection of two-level atoms can then be interpreted as a large
``spin'' of magnitude $j=N/2$.
For the photons we use the basis of Fock states
$\ket{n}$ with $n=0, 1, 2, 3, \ldots$, defined by
\be
  \op{a}^{\dagger}\ket{n}=\sqrt{n+1}\ket{n+1}, \quad
  \op{a}\ket{n}=\sqrt{n}\ket{n-1}.
\ee

The Dicke Hamiltonian $\op{H}_{\mathrm{D}}$ has an important symmetry,
it is invariant under a parity operation.
To see this, we introduce the unitary transformation
\be
  \op{U}(\varphi) = \exp\lb i \varphi \op{N}_{\mbox{\scriptsize{ex}}} \rb,
\ee
where the operator
$\op{N}_{\mbox{\scriptsize{ex}}}=\op{a}^{\dagger}\op{a} + \op{J}_{z} + j$ 
counts the number of excited quanta in the system.
In view of the relations
\be
  \op{U} \op{a} \op{U}^{\dagger} = e^{i\varphi} \op{a}, \quad
  \op{U} \op{J}_{+} \op{U}^{\dagger} = e^{i\varphi} \op{J}_{+}
\ee
the transformed Dicke Hamiltonian is given by
\begin{align}
  \label{eq:dickeham_transformed}
  & \op{U} \op{H}_{\mathrm{D}} \op{U}^{\dagger} 
  =
  \omega_{0} \op{J}_{z} + \omega \op{a}^{\dagger}\op{a} \nonumber \\
  &+ \frac{\lambda}{\sqrt{2j}}
   \lp 
    \op{a}^{\dagger}\op{J}_{+}e^{2i\varphi}
   + \op{a}^{\dagger}\op{J}_{-}
   + \op{a}\op{J}_{+}
   + \op{a}\op{J}_{-} e^{-2i\varphi} 
   \rp
\end{align}
and thus remains invariant if $\varphi$ is a multiple of $\pi$.
Therefore the Dicke model is symmetric under the parity transformation defined as 
\be
  \label{eq:paritydm}
   \op{\Pi}=\op{U}(\pi) = \exp\lb i \pi \op{N}_{\mbox{\scriptsize{ex}}} \rb.
\ee
The eigenvalues of $\op{\Pi}$ are $e^{i \pi (n+m+j)} = \pm 1$.
Due to this $\mathbb{Z}_{2}$ symmetry the Hilbert space breaks up into
two disjoint pieces which are not coupled by the Hamiltonian, one with
an even, the other with an odd excitation number $n+m+j$.

The Dicke model~(\ref{eq:dickeham}) has a quantum phase transition in
the thermodynamic limit $N \to \infty$ ($j \to \infty$) at the quantum
critical point $\lambda_{c}^0 = \sqrt{\omega\omega_{0}}/2$.
Both the mean photon number $\langle \op{a}^\dag \op{a}\rangle/j$ and
the atomic inversion $(\langle \op{J}_z\rangle +j)/j$ 
vanish for $\lambda<\lambda_c^0$ and are finite for
$\lambda>\lambda_c^0$. 
Numerical results for finite $j$ show that the thermodynamic limit is
approached very quickly.
Already the curves for $j=5$ follow the $j\rightarrow\infty$ results
rather closely, except in a small region around the critical
point~\cite{Emary2003}.

Eq.~(\ref{eq:dickeham_transformed}) shows that only the
counter-rotating terms are affected by the transformation
$\op{U}(\varphi)$. 
Within the rotating wave approximation, where these terms are
neglected, the Dicke Hamiltonian does not depend on the angle
$\varphi$ and therefore has ${\mathbb{U}}(1)$ symmetry.
Nevertheless there is again a quantum phase transition with a similar
behavior for the mean photon number (and also for the atomic
inversion), except that the critical coupling of the full Dicke model
is multiplied by a factor of $2$.

\section{Time-dependent Dicke model}
\label{sec:timedependentdickemodel}

\subsection{Model and model parameters}

\begin{figure}[h!]
  \begin{center}
    \includegraphics[scale=0.4]{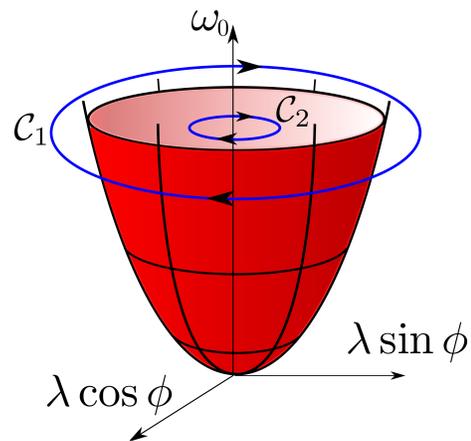}
    \caption{
      Critical surface of the Dicke model (for the case of a
      time-independent angle $\phi$) and circular paths chosen in the
      study of the time-dependent Dicke model. 
      }
    \label{fig:dickemodeleqphdiag}
  \end{center}
\end{figure}

The focus of our investigation is on a rotationally driven version of the Dicke
model~\cite{Plastina2006}.
The unitary transformation
\be
  \op{R}_{z}(t) = \exp\lb -i\phi(t)\op{J}_{z} \rb
  \label{rot_op}
\ee
generates a rotation of the spin $\op{\bf J}$ (of length $j=N/2$)
around the $z$ axis with a time-dependent angle $\phi(t)$.
Applying this transformation to the Dicke
Hamiltonian~(\ref{eq:dickeham}) we get 
\begin{align}
 \label{eq:hamrd}
& \op{H}_{\mathrm{RD}}(t)
= \op{R}_{z}^{\dagger}(t) \op{H}_{\mathrm{D}} \op{R}_{z}(t) \nonumber\\
&= \omega_{0} \op{J}_{z} +
   \omega \op{a}^{\dagger}\op{a} +
   \frac{\lambda}{\sqrt{2j}} \lp \op{a}^{\dagger} + \op{a} \rp \lp
   e^{i\phi(t)}\op{J}_{+} + e^{-i\phi(t)}\op{J}_{-} \rp.
\end{align}
We choose a linear time dependence 
\be
  \phi(t) = \delta_{\phi} \, t,
\ee
so that the two-dimensional vector $(\lambda\cos\phi,
\lambda\sin\phi)$ describes a circular motion with constant velocity
$\delta_\phi$.
This is illustrated in Fig.~\ref{fig:dickemodeleqphdiag}, where two
distinct paths are shown.
The path ${\cal C}_2$ lies inside the critical surface, defined by the equation 
$\omega_0=4\lambda^2/\omega$, and thus probes the normal phase, while
the path ${\cal C}_1$ lies outside and probes the superradiant phase,
where intriguing dynamical effects appear.

In the following the analysis will be developed for general values of
the parameters, but concrete calculations will be carried out at
resonance, and we will choose $\omega=\omega_0=1$.
In this case there remain two relevant parameters, the coupling strength 
$\lambda$ and the driving velocity $\delta_\phi$.

\subsection{Dynamics}

The rotationally driven Dicke Hamiltonian $\op{H}_{\mathrm{RD}}(t)$ is
obtained by applying a unitary transformation, therefore its
eigenvalues are the same as those of the original Hamiltonian
$\op{H}_{\mathrm{D}}$.
Nevertheless, the evolution, governed by the time-dependent
Schr\"odinger equation,
$i\partial_{t}\ket{\psi_{\mbox{\scriptsize RD}}(t)} =
\op{H}_{\mathrm{RD}}(t) \ket{\psi_{\mbox{\scriptsize RD}}(t)}$
($\hbar=1$), 
will lead to novel results, thanks to the combined effects of
geometric and dynamic phases.
Our approach is to start with a specific initial state at $t=0$ and to
let the system evolve according to the Schr\"odinger equation up to a
time $t>0$, where some quantities are measured.
A particularly important quantity is the mean photon number
\be
  n_{\mbox{\scriptsize ph}}(t)
  =
  \frac{1}{j} \,
  \bra{\psi_{\mbox{\scriptsize RD}}(t)} \op{a}^{\dagger}\op{a} \ket{\psi_{\mbox{\scriptsize RD}}(t)},
\label{photon_density}
\ee
which in the case of the time-independent Dicke Hamiltonian can be
taken as order parameter.
Alternatively one may also consider the atomic inversion
\be
  n_{\mbox{\scriptsize at}}(t)
  =
  1
  + 
  \frac{1}{j} \,
  \bra{\psi_{\mbox{\scriptsize RD}}(t)} \op{J}_z \ket{\psi_{\mbox{\scriptsize RD}}(t)}.
  \label{atomic_inversion}
\ee 
The sum of these two quantities is just the average number of excitations
\be
  n_{\mbox{\scriptsize ex}}(t)
  =
  \frac{1}{j} \,
  \bra{\psi_{\mbox{\scriptsize RD}}(t)} \op{N}_{\mbox{\scriptsize ex}} \ket{\psi_{\mbox{\scriptsize RD}}(t)}.
\ee 

The analysis is greatly facilitated by a transformation to a co-rotated state,
\be
\vert\psi_{\mbox{\scriptsize ROT}}(t)\rangle
=
\op{R}_z(t)\vert\psi_{\mbox{\scriptsize RD}}(t)\rangle,
\label{transf}
\ee
which evolves according to the Schr\"odinger equation
\be
i\partial_{t}\ket{\psi_{\mbox{\scriptsize ROT}}(t)} = (\op{H}_{\mathrm{D}}+\delta_\phi\op{J}_z)
\ket{\psi_{\mbox{\scriptsize ROT}}(t)} \, .
\label{schr_rot}
\ee
The operator
\be
\op{H}_{\mbox{\scriptsize ROT}}
=
\op{H}_{\mbox{\scriptsize D}}+\delta_\phi\, \op{J}_z
\label{rotated_ham}
\ee
is time-independent and has the form of the Dicke Hamiltonian, with
$\omega_0$ replaced by the renormalized level splitting
\be
\Omega = \omega_0+\delta_\phi. 
\label{renorm_split}
\ee
Therefore its ground state exhibits a quantum phase transition at
$\lambda_c=\frac{1}{2}\sqrt{\Omega\omega}$ from a normal phase with a
vanishing photon density to a superradiant phase with a finite photon
density.
The critical value $\lambda_c$ is expected to play a central role in
the time-dependence of physical quantities.

The expectation value of an operator $\op{O}$ which commutes with
$\op{J}_z$ is not affected by the transformation to a co-rotated
frame,
\be
[\op{O},\op{J}_z]=0 \Rightarrow\, 
\langle\psi_{\mbox{\scriptsize RD}}(t)\vert\op{O}\vert\psi_{\mbox{\scriptsize RD}}(t)\rangle=\, 
\langle\psi_{\mbox{\scriptsize ROT}}(t)\vert\op{O}\vert\psi_{\mbox{\scriptsize ROT}}(t)\rangle\, .
\label{exp_values}
\ee
This is the case for the quantities mentioned above.
If $\vert\psi_{\mbox{\scriptsize RD}}(0)\rangle$ happens to be an
eigenstate of $\op{H}_{\mbox{\scriptsize ROT}}$,
then Eq. (\ref{exp_values}) implies that  
$\langle\psi_{\mbox{\scriptsize RD}}(t)\vert\op{O}\vert\psi_{\mbox{\scriptsize RD}}(t)\rangle$ 
is time-independent.
Specifically, if $\vert\psi_{\mbox{\scriptsize RD}}(0)\rangle$ is the
ground state of $\op{H}_{\mbox{\scriptsize ROT}}$, both the photon
density (\ref{photon_density}) and the fraction of excited atoms
(\ref{atomic_inversion}) will be time-independent and equal to the
respective ground state expectation values for
$\op{H}_{\mbox{\scriptsize ROT}}$.

With coherent states, discussed in the next section, expectation
values of the operators $\op{a}$, $\op{a}^\dag$ and $\op{J}_{\pm}$ are
also relevant.
While for the photon creation and annihilation operators the
equality~(\ref{exp_values}) holds, this is no longer true in the case
of the operators $\op{J}_{\pm}$, for which we get
\be
\langle\psi_{\mbox{\scriptsize RD}}(t)\vert \op{J}_\pm \vert\psi_{\mbox{\scriptsize RD}}(t)\rangle
= 
e^{\pm i\phi(t)}
\langle\psi_{\mbox{\scriptsize ROT}}(t)\vert \op{J}_\pm \vert\psi_{\mbox{\scriptsize ROT}}(t)\rangle\, .
\label{rotation}
\ee
If $\vert\psi_{\mathrm{RD}}(0)\rangle$ is an eigenstate of
$\op{H}_{\mbox{\scriptsize ROT}}$ the vector
$\langle\psi_{\mbox{\scriptsize RD}}(t)\vert\op{\bf
  J}\vert\psi_{\mbox{\scriptsize RD}}(t)\rangle$ 
precesses with an angular velocity $\delta_\phi$ around the $z$ axis.

\subsection{Numerical procedure}
\label{sec_numerical}

We are mostly interested in the thermodynamic limit,
$N\rightarrow\infty$, where mean-field theory is expected to produce
exact results.
In the next Section we will explain in detail how to construct a
time-dependent mean-field theory using coherent states.
Calculations for a finite number of two-level atoms $N=2j$ 
(or for a pseudo-spin of finite length $j=N/2$) are nevertheless
useful because they allow us both to check the validity of mean-field
theory and to study the approach to the thermodynamic limit. 
In our numerical procedure we truncate the bosonic Hilbert space
up to $n_{\mathrm{M}}$ bosons but keep the full $(2j+1)$-dimensional
Hilbert space of the spin.
We choose $n_{\mathrm{M}}$ always high enough to assure that the error
of the numerical data is on the level of the machine precision.
In the co-rotating frame the evolution is determined by the
time-independent Hamiltonian $\op{H}_{\mbox{\scriptsize ROT}}$, 
Eq.~(\ref{rotated_ham}).
To calculate dynamic quantities~(\ref{exp_values}) we use the
Chebyshev scheme~\cite{Tal1984}, where the evolution operator for a
small time interval $\Delta t$ is expanded in a Chebyshev series,
\be
 e^{-i \op{H}_{\mbox{\scriptsize ROT}}\Delta t}
 \approx
 \sum_{k=0}^{M} a_{k} T_{k}(\op{h}),
\ee
where $T_{k}(\op{h})$ are the Chebyshev polynomials of order
$k$ and $\op{h}$ is a rescaled Hamiltonian
\be
 \op{h}
 =
 \frac{2\op{H}_{\mbox{\scriptsize ROT}} - (E_{\mathrm{Max}}+E_{\mathrm{Min}})\op{1}}{E_{\mathrm{Max}}-E_{\mathrm{Min}}}.
\ee
Here $E_{\mathrm{Max}}$ and $E_{\mathrm{Min}}$ are, respectively, the
largest and the smallest eigenvalues of $\op{H}_{\mbox{\scriptsize ROT}}$.
The expansion coefficients $a_{k}$ are determined by
\begin{align}
  a_{k}
  =&
  (-i)^{k} 
  \exp\lp - i \Delta t \frac{1}{2}
  (E_{\mathrm{Max}} + E_{\mathrm{Min}})
  \rp
  (2-\delta_{k,0}) \nonumber \\
  & J_{k}\left(\Delta t\frac{1}{2}(E_{\mathrm{Max}} - E_{\mathrm{Min}})\right),
\end{align}
where $\delta_{k,0}$ is the Kronecker symbol and $J_{k}(x)$ are the
Bessel functions of the first kind.

\section{Time-dependent mean-field theory}
\label{sec_mfa}

To solve the Schr\"odinger equation we seek a method which is simple
to treat and at the same time becomes exact in the thermodynamic
limit.
Mean-field theory is believed to reproduce exactly the main features
of the equilibrium quantum phase transition of the Dicke model for 
$j\rightarrow\infty$~\cite{castanos}, and therefore it should also
yield good results for the dynamics in this limit.
Mainly two methods have been used for establishing mean-field
equations, the coherent-state representation~\cite{Zhang1990} and the
Holstein-Primakoff transformation~\cite{Emary2003}.
These two approaches can be shown to be equivalent in the
thermodynamic limit~\cite{kapor}.
We adopt the coherent-state representation, using a similar procedure
as the one presented in Ref.~\cite{aguiar}.

\subsection{Coherent-state representation}

We have seen that for the relevant operators $\op{O}$ the expectation
value
$\bra{\psi_{\mbox{\scriptsize RD}}(t)} \op{O} \ket{\psi_{\mbox{\scriptsize RD}}(t)}$
with an evolution governed by the time-dependent Hamiltonian
$\op{H}_{\mbox{\scriptsize RD}}(t)$ is simply related to
$\langle\psi_{\mbox{\scriptsize ROT}}(t)\vert\op{O}\vert\psi_{\mbox{\scriptsize ROT}}(t)\rangle$ 
with
$\vert\psi_{\mbox{\scriptsize ROT}}(t)\rangle$ evolving according to
the time-independent Hamiltonian $\op{H}_{\mbox{\scriptsize ROT}}$, 
Eq.~(\ref{rotated_ham}).
For operators commuting with $\op{J}_z$ the two expectation values are
the same, while they are transformed to each other by the rotation
(\ref{rotation}) for the operators $\op{J}_\pm$.
In the following we calculate expectation values with respect to
co-rotated states and transform the resulting expressions back to the
unrotated frame at the end, if necessary.
We use the notation
\be
\langle\op{O}\rangle(t)
:=\langle\psi_{\mbox{\scriptsize ROT}}(t)\vert\op{O}\vert\psi_{\mbox{\scriptsize ROT}}(t)\rangle\, .
\ee
The Schr\"odinger equation (\ref{schr_rot}) leads to
\be
\partial_t\langle\op{O}\rangle(t)=i\langle[\op{H}_{\mbox{\scriptsize ROT}},\op{O}]\rangle(t)\, .
\ee
Specifically we find the equations of motion
\begin{align}
  \partial_t\langle\op{J}_z\rangle(t)
  &=
  \frac{i\lambda}{\sqrt{2j}}
  \langle(\op{a}+\op{a}^\dag)(\op{J}_--\op{J}_+)\rangle(t)\, ,\nonumber\\
  \partial_t\langle\op{J}_-\rangle(t)
  &=
  -i (\omega_0+\delta_\phi)\langle\op{J}_{-}\rangle(t)
  +i \sqrt{\frac{2}{j}}\lambda\langle (\op{a}+\op{a}^\dag)\op{J}_z\rangle(t)\, ,\nonumber\\
  \partial_t\langle\op{a}\rangle(t)
  &=
  -i\omega\langle\op{a}\rangle(t)
  -i\frac{\lambda}{\sqrt{2j}}\langle(\op{J}_-+\op{J}_+)\rangle(t)\, .
\label{eqmotion1}
\end{align}
Those for $\langle\op{J}_+\rangle(t)$ and
$\langle\op{a}^\dag\rangle(t)$ are obtained by complex conjugation.

Our main assumption is now that $\vert\psi_{\mbox{\scriptsize ROT}}(t)\rangle$ is a coherent
state, not only for $t=0$, but also for $t>0$. 
In the present context this means
\be
\vert\psi_{\mbox{\scriptsize ROT}}(t)\rangle=\vert\alpha(t)\rangle\otimes\vert\zeta(t)\rangle
=:
\vert\alpha(t),\zeta(t)\rangle
\label{cohst}
\ee
with spin-coherent states
\be
  \vert\zeta(t)\rangle
  =
  (1 + \vert\zeta(t)\vert^2)^{-j}\, e^{\zeta(t)\op{J}_{+}}\vert j,-j\rangle
\ee
and bosonic coherent states
\be
  \vert\alpha(t)\rangle
  = 
  e^{- \frac{1}{2} \vert\alpha(t)\vert^2}\, e^{\alpha(t)a^{\dagger}}\vert 0\rangle.
\ee
The expectation values of the relevant operators with respect to the
coherent state~(\ref{cohst}) are
\begin{align}
  \langle \op{J}_{+} \rangle(t)
  &= 
  2j \frac{\zeta^*(t)}{1+\vert\zeta(t)\vert^2}
  =
  \langle\op{J}_{-}\rangle^*(t)\, , \nonumber \\
  \langle\op{J}_{z}\rangle(t)
  &=
  -j \, \frac{1-\vert\zeta(t)\vert^2}{1+\vert\zeta(t)\vert^2}\, ,\nonumber \\
  \langle\op{a}\rangle(t)
  &=
  \alpha(t)\, ,\nonumber\\
  \langle\op{a}^\dag\op{a}\rangle(t)
  &=
  \vert\alpha(t)\vert^2\, .
  \label{expval}
\end{align}
We note that if $\vert\psi_{\mbox{\scriptsize ROT}}(t)\rangle$ is a
coherent state then this is also true for
$\vert\psi_{\mbox{\scriptsize RD}}(t)\rangle$, which is related to
$\vert\psi_{\mbox{\scriptsize ROT}}(t)\rangle$ by Eq.~(\ref{transf}),
and vice versa.
To see this one verifies that the application of 
$\op{R}_z^\dag(t)$ on $\vert\zeta(t)\rangle$
produces a spin-coherent state $\vert\zeta_{\mbox{\scriptsize RD}}(t)\rangle$
in the original frame with
\be
\zeta_{\mbox{\scriptsize RD}}(t)
=
e^{i\phi(t)}\zeta(t).
\label{transf_coh}
\ee

\subsection{Classical dynamical system}
\label{subsec_classical}

The equations of motion (\ref{eqmotion1}) can now be linked to a
classical dynamical system of two coupled oscillators using the
parametrization
\beq
\zeta(t)
&=&
\frac{1}{\sqrt{4-[Q^2(t)+P^2(t)]}} \, [Q(t) + i P(t)], \nonumber\\
\alpha(t)
&=&
\sqrt{\frac{j}{2}}\, [q(t)+ip(t)].
\label{def_coord}
\eeq
Inserting these relations into Eqs.~(\ref{eqmotion1}) and isolating
real and imaginary parts we find equations of motion 
\beq
&\dot{Q}&=\Omega P-\lambda q\frac{QP}{\sqrt{4-(Q^2+P^2)}}\, ,\nonumber\\
&\dot{P}&=-\Omega Q+\lambda q\frac{2Q^2+P^2-4}{\sqrt{4-(Q^2+P^2)}}\, ,\nonumber\\
&\dot{q}&=\omega p\nonumber\\
&\dot{p}&=-\omega q-\lambda Q\sqrt{4-(Q^2+P^2)}\, ,
\label{eqmotion2}
\eeq
where $\Omega$ is the renormalized level splitting~(\ref{renorm_split}).
With the same parametrization the expectation value of 
$\op{H}_{\mbox{\scriptsize ROT}}$ yields a classical Hamiltonian 
$H_{\mbox{\scriptsize cl}} := \Omega 
+\frac{1}{j}\langle\alpha(t),\zeta(t)\vert\op{H}_{\mbox{\scriptsize ROT}}\vert\alpha(t),\zeta(t)\rangle$, explicitly 
\be
 H_{\mbox{\scriptsize cl}}
 =
 \frac{\Omega}{2}(Q^2+P^2)
 +\frac{\omega}{2}(q^2+p^2)
 +\lambda Qq\sqrt{4-(Q^2+P^2)}.
 \label{ham_class}
\ee
Similarly, using Eq.~(\ref{expval}) we find for the mean photon number~(\ref{photon_density})
\be
 n_{\mbox{\scriptsize ph}}=\frac{1}{j}\vert\alpha\vert^2=\frac{1}{2}(q^2+p^2)\, ,
\ee
while the atomic inversion (\ref{atomic_inversion}) is found to be
\be
 n_{\mbox{\scriptsize at}}=\frac{2\vert\zeta\vert^2}{1+\vert\zeta\vert^2}
 =\frac{1}{2}(Q^2+P^2)\, .
\ee
One readily verifies that Eqs.~(\ref{eqmotion2}) are the canonical
equations for the Hamiltonian~(\ref{ham_class}) and therefore
$Q,P,q,p$ are the canonical coordinates of a dynamical system of two
degrees of freedom. 
Once the equations~(\ref{eqmotion2}) have been solved it is easy to
find the corresponding coordinates in the original system using
Eq.~(\ref{transf_coh}).
The photonic variables $q,p$ remain the same, while the atomic variables $Q,P$ become
\beq
Q_{\mbox{\scriptsize RD}}(t)&=&Q(t)\cos\phi(t)-P(t)\sin\phi(t)\, ,\nonumber\\
P_{\mbox{\scriptsize RD}}(t)&=&Q(t)\sin\phi(t)+P(t)\cos\phi(t)
\label{transf_can}
\eeq
in the unrotated frame.

We have verified that the same classical Hamiltonian emerges either
when applying the Holstein-Primakoff transformation from spins to
bosons or when using the saddle-point approximation of the
coherent-state path integral~\cite{Zhang1990}.

\subsection{Stationary states and linear modes}
\label{subsec_linear}

We show now that the stationary solutions of the classical equations
of motion reproduce the phases characterizing the ground state of
$\op{H}_{\mbox{\scriptsize ROT}}$ in the limit $j\rightarrow\infty$.
The arguments are similar to those used in Ref.~\cite{Emary2003} for
the time-independent Dicke Hamiltonian.
Inserting $\dot{Q}=\dot{P}=\dot{q}=\dot{p}=0$ into
Eqs.~(\ref{eqmotion2}) we find a first obvious solution
$Q_0=P_0=q_0=p_0=0$ for
$\lambda<\lambda_c=\frac{1}{2}\sqrt{\omega\Omega}$.
It describes the normal phase with vanishing photon density.
The linearized equations of motion around this point are
\beq
&\dot{Q}&=\Omega P\, ,\nonumber\\
&\dot{P}&=-\Omega Q-2\lambda q\, ,\nonumber\\
&\dot{q}&=\omega p\nonumber\, ,\\
&\dot{p}&=-\omega q-2\lambda Q\, ,
\label{eqsmot1}
\eeq
and the eigenmode frequencies are given by
\be
\varepsilon_{\pm}^2=\frac{1}{2}\left\{\Omega^2+\omega^2
\pm\sqrt{(\Omega^2-\omega^2)^2+16\lambda^2\Omega\omega}\right\}\, .
\label{eigenfr1}
\ee
$\varepsilon_{-}$ tends to zero for $\lambda\rightarrow\lambda_c$ and
thus represents the soft mode of the transition.

Two stationary solutions are found for $\lambda>\lambda_c$, namely 
\beq
Q_0&=&\pm \sqrt{2}\left(1-\frac{{\lambda_c}^2}{\lambda^2}\right)^{\frac{1}{2}}\, ,\nonumber\\
q_0&=&\mp\frac{2\lambda}{\omega}\left(1-\frac{{\lambda_c}^4}{\lambda^4}\right)^\frac{1}{2}\, ,
\nonumber\\
P_0&=&p_0\, =\, 0\, .
\label{stationary}
\eeq 
They describe the superradiant phase with densities
\beq
 n_{\mbox{\scriptsize ph}}
 &=&
 \frac{1}{j}\vert\alpha\vert^2
 =
 \frac{2\lambda^2}{\omega^2}\left(1-\frac{{\lambda_c}^4}{\lambda^4}\right)\, ,\nonumber\\
 n_{\mbox{\scriptsize at}}
 &=&
 \frac{2\vert\zeta\vert^2}{1+\vert\zeta\vert^2}
 = 1-\frac{{\lambda_c}^2}{\lambda^2}\, .
 \label{eq:densities_mfa}
\eeq

We consider small deviations $X=Q-Q_0$, $x=q-q_0$, $P$, $p$ from the
stationary solutions~(\ref{stationary}) of the superradiant phase.
To linear order in these coordinates the equations of
motion~(\ref{eqmotion2}) read
\beq
&\dot{X}&=\frac{2}{\omega}(\lambda^2+{\lambda_c}^2)P\, ,\nonumber\\
&\dot{P}&=-\frac{\sqrt{8}{\lambda_c}^2}{\sqrt{\lambda^2+{\lambda_c}^2}}x
-\frac{8\lambda^4}{\omega(\lambda^2+{\lambda_c}^2)}X\, ,\nonumber\\
&\dot{x}&=\omega p\, ,\nonumber\\
&\dot{p}&=-\omega x-\frac{\sqrt{8}{\lambda_c}^2}{\sqrt{\lambda^2+{\lambda_c}^2}}X\, .
\label{eqsmot2}
\eeq
The eigenmode frequencies of this linear system are
\be
\varepsilon_{\pm}^2=\frac{1}{2}\left\{\omega^2+\frac{16\lambda^4}{\omega^2}
\pm\sqrt{\left(\omega^2-\frac{16\lambda^4}{\omega^2}\right)^2+4(\omega\Omega)^2}\right\}\, .
\label{eigenfr2}
\ee
The solution with eigenfrequency $\varepsilon_{-}$ again represents a soft mode, 
$\varepsilon_{-} \rightarrow 0$ for $\lambda\rightarrow\lambda_c$. 
Eqs.~(\ref{eigenfr1}), (\ref{stationary}) and (\ref{eigenfr2}) agree
with the corresponding expressions for the undriven Dicke model in
Ref.~\cite{Emary2003}, where the Holstein-Primakoff transformation has
been used.

This stability analysis shows that in mean-field approximation the
rotation applied to the Dicke Hamiltonian changes the phase diagram.
The critical point $\lambda_c^0=\sqrt{\omega\omega_0}/2$ is shifted to
$\lambda_c=\sqrt{\omega\Omega}/2$, where
$\Omega=\omega_0+\delta_\phi$.
The expressions for the photon density and the atomic inversion remain
the same as in the time-independent Dicke model, except that 
$\lambda_c^0$ is replaced by $\lambda_c$. Also the eigenmode frequencies of the
time-dependent Dicke Hamiltonian are simply obtained from those of the
time-independent Hamiltonian by replacing $\omega_0$ by $\Omega$.
\begin{figure}[h!]
  \begin{center}
  \includegraphics[scale=0.6]{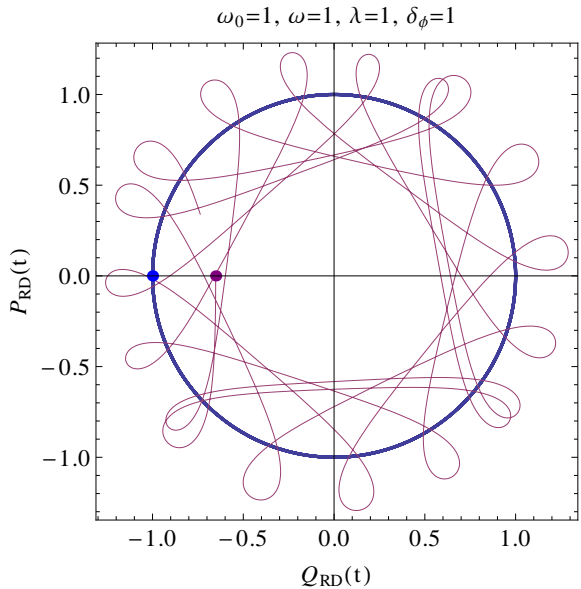}
    \caption{
    Phase portrait
    $(Q_{\mbox{\scriptsize{RD}}}(t),P_{\mbox{\scriptsize RD}}(t))$:
    The blue circle represents the stationary solution in the
    co-rotating frame, corresponding
    to the initial condition $Q_{\mbox{\scriptsize{RD}}}(0)=Q_{0},
    \, P_{\mbox{\scriptsize{RD}}}(0)=0, \, q(0)=q_{0}, \, p(0)=0$. 
    The purple trajectory has the initial condition slightly away from
    the stationary point, i.e., $Q_{\mbox{\scriptsize{RD}}}(0)=Q_{0} +
    0.347851, \, P_{\mbox{\scriptsize{RD}}}(0)=0, \, q(0)=q_{0}, \, p(0)=0$.}
    \label{fig:phaseportrait}
  \end{center}
\end{figure}

The stationary points for the co-rotating frame are readily
transformed to the original frame using Eq.~(\ref{transf_can}).
The stationary point of the normal phase does not change, but that of
the superradiant phase is transformed into uniform circular motion in
the $Q_{\mbox{\scriptsize RD}}$-$P_{\mbox{\scriptsize RD}}$ plane with radius 
$2\sqrt{n_{\mbox{\scriptsize at}}}$ and angular velocity $\delta_\phi$.

For an initial condition which is close to a stationary point (in the
co-rotated frame) the evolution of the system is governed by the
linearized equations of motion and will in general be a superposition
of harmonic oscillations with frequencies $\varepsilon_{\pm}$. 
The trajectories form complicated Lissajous patterns, as exemplified
in Fig.~\ref{fig:phaseportrait}.
The purple trajectory (thin line) starting close to the stationary
circle (thick line) shows that in addition to the circular motion of
angular frequency $\delta_{\phi}$ there is a transverse harmonic
oscillation with a frequency 
$\epsilon_{-} \approx 1 - \frac{1}{8\lambda^{4}} = 0.875$ (see
Appendix~\ref{A}).

\section{Dynamically generated darkness}
\label{darkness}

We discuss now the evolution after the rotation (\ref{rot_op}) is
switched on suddenly at $t=0$.
Thus the system is supposed to be in the ground state
$\vert\psi_0\rangle$ of the time-independent Dicke Hamiltonian
$\op{H}_{\mbox{\scriptsize D}}$ for $t<0$ and to evolve according to
the time-dependent Dicke Hamiltonian $\op{H}_{\mbox{\scriptsize RD}}(t)$
for $t>0$.
In the rotated frame we have to solve the Schr\"odinger equation
(\ref{schr_rot}) for $\vert\psi_{\mbox{\scriptsize ROT}}(t)\rangle$
with the initial condition 
$\vert\psi_{\mbox{\scriptsize ROT}}(0)\rangle=\vert\psi_0\rangle$.
We have used both the numerical technique for the full quantum problem 
(explained in Section~\ref{sec_numerical}) and the mean-field
approximation (described in Section~\ref{sec_mfa}) for calculating
important quantities such as the photon number and the atomic
inversion. 

Nothing special will happen for $\lambda<\lambda_c^0$, where the
system starts in the normal phase of $\op{H}_{\mbox{\scriptsize D}}$
for $t=0$ and proceeds in the normal phase of
$\op{H}_{\mbox{\scriptsize RD}}(t)$ for $t>0$. 
Therefore we concentrate on couplings $\lambda>\lambda_c^0$.
We use preferentially parameter values
$\omega=\omega_0=\delta_\phi=1$, for which the critical points are
$\lambda_c^0=0.5$ and $\lambda_c=1/\sqrt{2}\approx 0.707$.
For the results presented here we have chosen as initial state the
mean-field ground state
of $\op{H}_{\mbox{\scriptsize D}}$, both for the numerical evaluation
of the exact quantum dynamics and for the treatment of the classical
dynamical system. 
We have verified that starting with the exact ground state of 
$\op{H}_{\mbox{\scriptsize D}}$ (for the numerical procedure) does not
modify the main results.
The initial values of the classical coordinates $Q(0)$, $q(0)$, $P(0)$ and $p(0)$ 
are given by Eqs.~(\ref{stationary}) with $\lambda_c$ replaced by
$\lambda_c^0$.
In view of Eqs.~(\ref{def_coord}) the corresponding parameters of the
initial coherent state are 
\beq
 \alpha(0)
 &=&
 \sqrt{\frac{j}{2}}q(0)
 =
 \mp \frac{\sqrt{2j}}{\omega\lambda}\sqrt{\lambda^4-[\lambda_c^0]^4}\, ,\nonumber\\
 \zeta(0)
 &=&
 \frac{Q(0)}{\sqrt{4-Q^2(0)}}
 =
 \pm\sqrt{\frac{\lambda^2-[\lambda_c^0]^2}{\lambda^2+[\lambda_c^0]^2}}\, .
\eeq
\begin{figure}[h!]
  \begin{center}
    \includegraphics[scale=0.43]{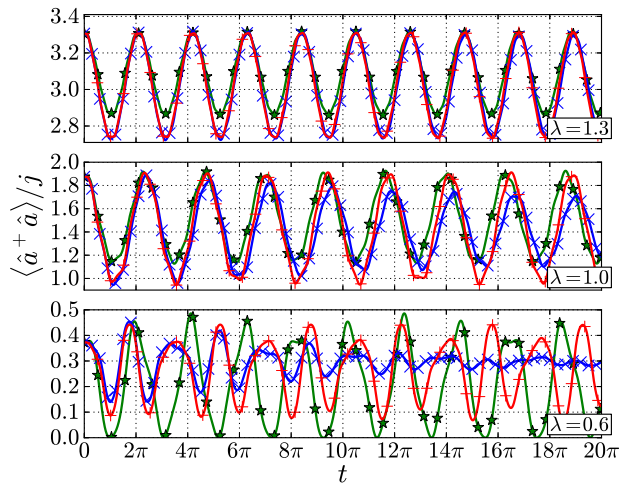}
    \caption{
      Time dependence of the mean photon number at resonance
      ($\omega = \omega_{0} = 1.0$) and for a driving velocity
      $\delta_{\phi}=1.0$. 
      The red curves (symbols $+$) represent the time evolution of the
      nonlinear mean-field equations, the green curves (symbols $*$)
      show the linearized limit of these equations, while the blue curves (symbols $\times$)
      have been obtained by numerical integration of the Schr\"odinger
      equation for a finite system ($j=10$, \textbf{$n_{\mathrm{M}}=60$}).
      The first two cases (couplings $\lambda=1.3,\, 1$) correspond to
      the superradiant phase of the time-dependent Dicke model, the
      third ($\lambda=0.6$) to its normal phase. All three
      couplings exceed the critical value of the time-independent
      Dicke model ($\lambda_c^0=0.5$).
     }
    \label{fig:adatimescs}
  \end{center}
\end{figure}

Fig.~\ref{fig:adatimescs} shows results for the mean photon number for
three different couplings.
Deep in the superradiant phase of $\op{H}_{\mbox{\scriptsize RD}}$
($\lambda=1.3$) a simple oscillatory behavior is observed, for both
mean-field and full quantum solutions, with $j=10$ in the latter case.
This behavior is readily understood in the linearized limit of the
mean-field equations, where the coordinates execute harmonic motions,
with frequencies $\varepsilon_{\pm}$, as explained in
Appendix~\ref{A}.
For $\lambda \gg \lambda_c$ the main contribution comes from the mode
with frequency $\varepsilon_{-} \approx 1-1/{8\lambda^4}$, in perfect
agreement with Fig.~\ref{fig:adatimescs}.
The results for $\lambda=1$ are still reproduced rather well by the
linearized dynamics, but not for $\lambda=0.6$, where the anharmonic
part of the potential becomes important for the chosen initial
conditions.
For this coupling the (nonlinear) mean-field behavior agrees only
initially with the dynamics obtained from the numerical integration of
the Schr\"odinger equation, but not at later times where the quantum
evolution appears to reach a constant value.
We attribute this difference to decoherence in the quantum case, for
which at longer times than those shown in the figure the initial
pattern is restored.
Mean-field theory remains by construction a coherent state. 

\begin{figure}[h!]
  \begin{center}
   \includegraphics[scale=0.45]{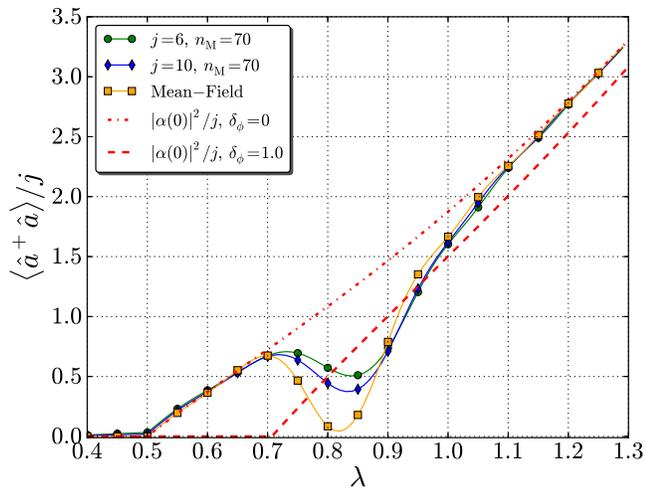}
    \caption{
      Mean photon number as a function of the
      coupling strength after a single period of revolution $T_\phi$
      and for $\omega=\omega_{0}=\delta_{\phi}=1$. 
      Dashed lines represent the ground state values for
      $\op{H}_{\mbox{\scriptsize D}}$ and $\op{H}_{\mbox{\scriptsize ROT}}$, respectively. 
      Full lines have been obtained starting with the ground state of
      $\op{H}_{\mbox{\scriptsize D}}$ at $t=0$, but evolving with
      $\op{H}_{\mbox{\scriptsize RD}}$ for $0<t<T_\phi$, both in
      mean-field approximation and by numerically solving the
      Schr\"odinger equation for finite $j$.
      }
    \label{fig:adalamscs}
  \end{center}
\end{figure}

We discuss now the mean photon number $n_{\mbox{\scriptsize ph}}(t)$.
If the system evolves according to $\op{H}_{\mbox{\scriptsize D}}$
and if the initial state $\vert\psi_{\mathrm{RD}}(0)\rangle$ is the ground state of
$\op{H}_{\mbox{\scriptsize D}}$, then $n_{\mbox{\scriptsize ph}}$ does
not depend on time and is just the order parameter of the
time-independent Dicke model, except that in the mean-field expression
Eq.~(\ref{eq:densities_mfa}) $\lambda_c$ is replaced by $\lambda_c^0$. 
An analogous statement holds if the evolution is governed by
$\op{H}_{\mbox{\scriptsize RD}}$ with the ground state of
$\op{H}_{\mbox{\scriptsize ROT}}$ as initial state, then
$n_{\mbox{\scriptsize ph}}$ is given by Eq.~(\ref{eq:densities_mfa}).
The two cases are illustrated as dashed lines in
Fig.~\ref{fig:adalamscs}.
A qualitatively different behavior is found for the scenario discussed
above, where the evolution is governed by $\op{H}_{\mbox{\scriptsize RD}}$, 
but the initial state is the ground state of 
$\op{H}_{\mbox{\scriptsize D}}$.
The full lines in Fig.~\ref{fig:adalamscs} show $n_{\mbox{\scriptsize ph}}$ 
after a single period of rotation $T_\phi:=2\pi/\delta_\phi$. 
For both $\lambda_c^0<\lambda<\lambda_c$ and
$\lambda\gg\lambda_c$ the mean photon number $n_{\mbox{\scriptsize ph}}(T_\phi)$ 
does not differ much from its initial value. 
For $\lambda$ slightly larger than $\lambda_c$ 
$n_{\mbox{\scriptsize ph}}(T_\phi)$ is strongly reduced as compared to
its value at $t=0$. 
This remarkable dynamically generated darkness is found both in
mean-field approximation and in the quantum evolution for finite $j$. 
\begin{figure}[h!]
  \begin{center}
    \includegraphics[scale=0.43]{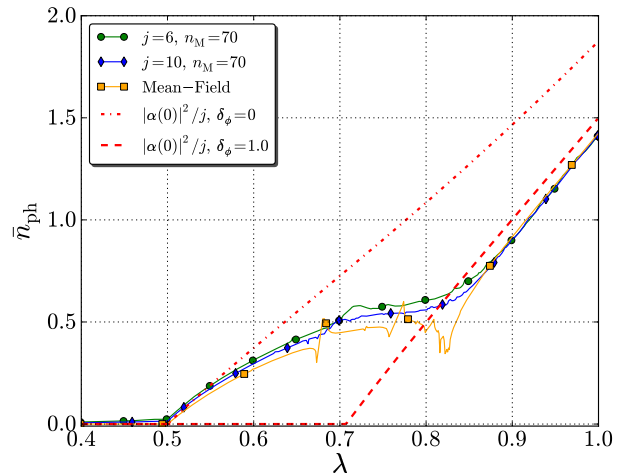}
    \caption{
      Coupling dependence of the mean photon number averaged over a
      time $T=150 \, T_\phi$ for parameter values
      $\omega=\omega_0=\delta_\phi=1$.       
      Dashed lines represent the ground state values for
      $\op{H}_{\mbox{\scriptsize D}}$ and $\op{H}_{\mbox{\scriptsize ROT}}$, respectively. 
      The green line (dots) has been obtained by numerically solving
      the Schr\"odinger equation for $j=6$ and the blue line
      (diamonds) for $j=10$ with a cutoff $n_{\mathrm{M}}=70$ in both cases.
      }
    \label{fig:adalamtascs}
  \end{center}
\end{figure}

The time-averaged mean photon number
\be
  \bar{n}_{\mbox{\scriptsize ph}}
  :=
  \frac{1}{jT}\int_0^{T}dt\, 
  \langle\psi(t)\vert\op{a}^\dag\op{a}\vert\psi(t)\rangle\,
\ee
is shown in Fig.~\ref{fig:adalamtascs} for $T \gg T_\phi$.
In contrast to $n_{\mbox{\scriptsize ph}}(T_\phi)$, 
$\bar{n}_{\mbox{\scriptsize ph}}$ approaches the steady state value of 
$\op{H}_{\mbox{\scriptsize RD}}$ for $\lambda\gg\lambda_c$ (and not that of 
$\op{H}_{\mbox{\scriptsize D}}$).
This is readily understood in the linearized limit of the classical
dynamics, where $q(t)$, $p(t)$ oscillate with a frequency 
$\varepsilon_{-}\approx\omega$ about the stationary values $q_0$,
$p_0$, as shown in Appendix~\ref{A}.
This implies that for $\omega=\delta_\phi=1$ the coordinates and thus
also $n_{\mbox{\scriptsize ph}}$ recover their original values after a
time $T_\phi$, in agreement with Fig.~\ref{fig:adalamscs}, but the
time-averaged mean photon number $\bar{n}_{\mbox{\scriptsize ph}}$ is
close to its stationary value, as in Fig.~\ref{fig:adalamtascs}. 
\begin{figure}[h!]
  \begin{center}
    \includegraphics[scale=0.43]{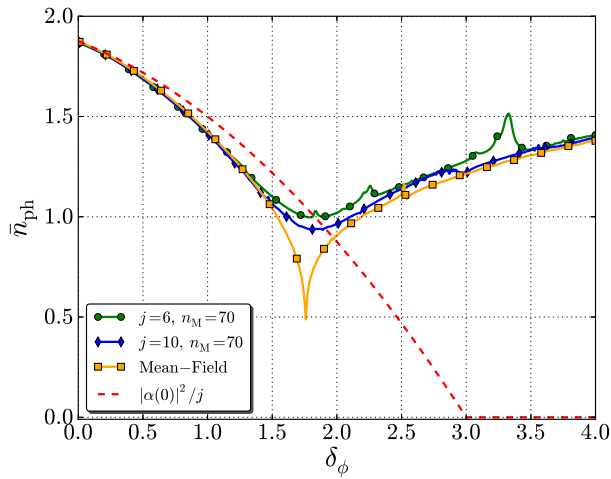}
    \caption{
      Time-averaged mean photon number (full lines), compared to the
      initial value (dashed line) as a function of the rotation
      velocity $\delta_\phi$ for $\lambda=\omega=\omega_0=1$ and 
      $T=150\, T_\phi$.
      There is a clear minimum, which is very sharp in mean-field
      approximation (yellow line) and more shallow when calculated
      numerically for the quantum case with $j=6$ and $j=10$ (green
      and blue lines, respectively).}
    \label{fig:adadphitacs}
  \end{center}
\end{figure}

Fig.~\ref{fig:adalamtascs} also shows that the reduction of the photon
number not only appears at $T_\phi$, but is also seen after taking the
time average. 
A clear minimum is found if the time-averaged photon number is plotted
as a function of the driving velocity $\delta_\phi$ for a fixed
coupling strength.
This is shown in Fig.~\ref{fig:adadphitacs}, where the darkening
effect is particularly pronounced in mean-field approximation.

To illustrate the phenomenon of dynamically generated darkness
we have studied a simple classical system of a particle moving on a
plane in an isotropic potential $V(\vert{\bf x}\vert)$ with a maximum
$V_{0}$ at the origin and a minimum at $\vert{\bf x}\vert=\rho_{0}$
(Mexican hat).
Details are described in Appendix~\ref{B}.
For a vanishing initial momentum ${\bf p}(0)$ and finite ${\bf x}(0)$
the particle exhibits radial oscillations around $\rho_0$ if the energy
$E$ is smaller than $V_0$, for $E>V_{0}$ it passes over the potential
hump at the center and moves back and forth between the turning points
$\pm {\bf x}_{\mathrm{max}}$. 
For $E \approx V_0$ the motion is slowed down close to the center,
where the particle spends most of the time.
The time average of both ${\bf x}^2(t)$ and ${\bf p}^2(t)$ tends to
zero for $E \rightarrow V_0$.
\begin{figure}[h!]
  \begin{center}
    \includegraphics[scale=0.41]{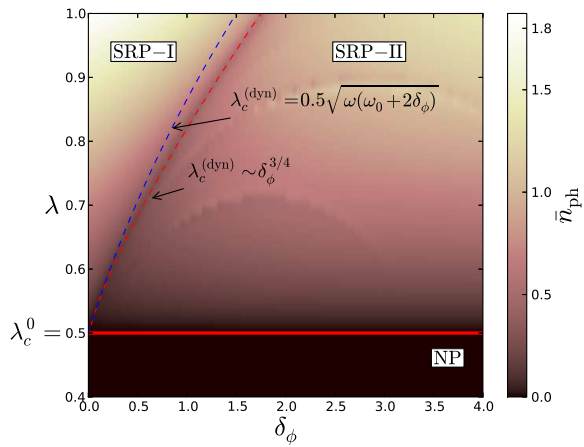}
    \caption{
      Time-averaged mean photon number as a function of both driving velocity
      $\delta_\phi$ and coupling strength $\lambda$, according to
      mean-field theory. 
      We have chosen parameter values $\omega=\omega_{0}=1.0$ and averaged over
      a time $T=150\, T_{\phi}$.
      The blue dashed line corresponds to the theoretical estimate of
      Eq.~(\ref{eq:dyncrit_anal}), the red dashed line is a fit to the
      observed minimum.
      }
    \label{fig:adaneqpdscs}
  \end{center}
\end{figure}

We use now the result found in the case of the Mexican hat to estimate
the critical coupling for maximal slowing down in the case of the
time-dependent Dicke model.
We expect this to occur if the energy for the given initial conditions
is equal to the maximum of the potential at the origin,
i.e. $H_{\mbox{\scriptsize cl}}$, given by Eq. (\ref{ham_class}),
vanishes for the coordinates of the stationary state of
$\op{H}_{\mathrm{D}}$.
We find the following estimate for this dynamic critical point
\be
\lambda_c^{\mbox{\scriptsize (dyn)}}\approx \frac{1}{2}\sqrt{\omega(\omega_0+2\delta_\phi)}\, .
\label{eq:dyncrit_anal}
\ee
This represents a rough guess because the potential in the case of the
Dicke model is both anisotropic and velocity dependent.
Therefore it is not surprising that the
expression~(\ref{eq:dyncrit_anal})
does not reproduce exactly the dynamic critical line found by solving
the mean-field equations and shown in Fig.~\ref{fig:adaneqpdscs}.
Rather the numerical data are well fitted by a power law, 
$\lambda_c^{\mbox{\scriptsize (dyn)}}=\lambda_c^0+0.327\, \delta_\phi^{3/4}$.

\begin{figure}[h!]
  \begin{center}
    \includegraphics[scale=0.63]{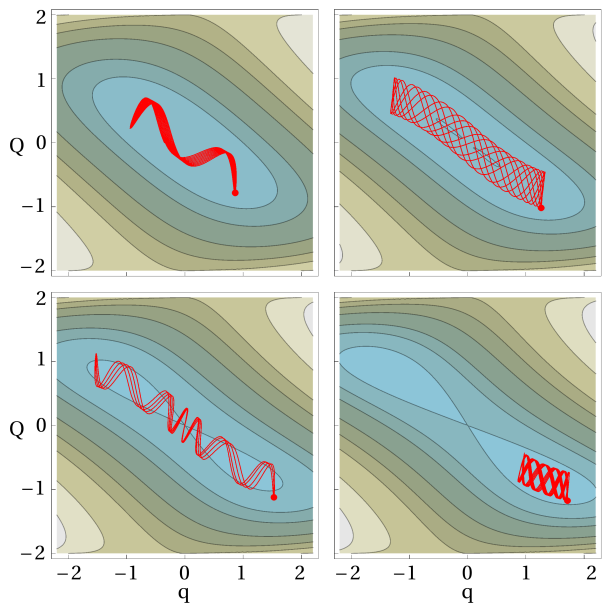}
    \caption{
      Contours of the potential $V_{\mathrm{cl}}(Q,q,0)$
      with the trajectories $(Q,q)$ illustrated by
      red curves. 
      The red dots mark the initial state given by
      Eqs.~(\ref{stationary}) with $\lambda_c$ replaced by
      $\lambda_c^0$.
      The four trajectories have been calculated for fixed parameter
      values $\omega=\omega_{0}=\delta_{\phi}=1$, but different
      couplings $\lambda=0.6, \, 0.72, \, 0.823, \, 0.9$
      (from left to right and top to bottom). 
      }
    \label{fig:effpot}
  \end{center}
\end{figure}

To exemplify the complicated dynamics of the time-dependent Dicke
model we present in Fig.~\ref{fig:effpot} a few trajectories, as
obtained by solving numerically the equations of
motion~(\ref{eqmotion2}).
The stationary states of the time-independent Dicke model have again
been chosen as initial conditions.
The contour lines in Fig.~\ref{fig:effpot} represent the static
limit $V_{\mathrm{cl}}(Q,q,0)$ of the classical potential
\be
  V_{\mathrm{cl}}(Q,q,P) 
  = 
  \frac{\Omega}{2}Q^{2}+\frac{\omega}{2}q^{2}+\lambda Q q \sqrt{4-(Q^{2}+P^{2})}.
\ee
The first trajectory is representative for the range
$\lambda_{c}^{0}<\lambda<\lambda_{c}$, where the motion is rather
regular,
while the fourth trajectory illustrates the strong-coupling region $\lambda \gg
\lambda_{c}$, where the motion is limited to the region around one
potential minimum.
This harmonic motion is correctly reproduced by the linearized limit
of the equation of motion.
The second trajectory shows the chaotic motion slightly above the
critical point ($\lambda=\lambda_{c}$).
The third example corresponds to the coupling constant for which
$\bar{n}_{\mathrm{ph}}$ has a pronounced minimum according to the
mean-field theory ($\lambda=0.823$, see Fig.~\ref{fig:adalamtascs}).
In this case the trajectory is seen to have inversion symmetry, thus
it passes through the origin.
However due to the transverse oscillations the motion never stops and
therefore the averaged mean photon number is only reduced but does not
tend to zero.

\begin{figure}[h!]
  \begin{center}
    \includegraphics[scale=0.44]{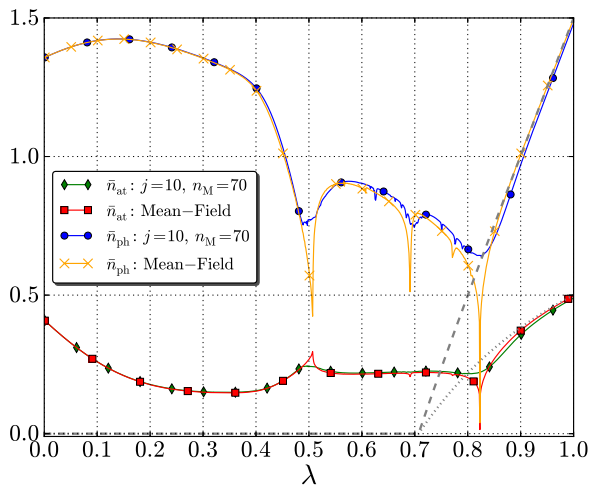}
    \caption{
      Coupling dependence of the mean photon number and the atomic
      inversion averaged over a time $T=150 T_{\phi}$ for
      $\omega=\omega_{0}=\delta_{\phi}=1$.
      Dashed lines represent the ground state values for
      $\op{H}_{\mathrm{ROT}}$.
      The blue line (dots) $\bar{n}_{\mathrm{ph}}$ and green line
      (diamonds) $\bar{n}_{\mathrm{at}}$ have been calculated by
      numerically solving the Schr\"odinger equation for $j=10$ and
      $n_{\mathrm{M}}=70$. 
      The mean field results, orange line (crosses) for
      $\bar{n}_{\mathrm{ph}}$ and red line (squares) for
      $\bar{n}_{\mathrm{at}}$, exhibit sharp features and drop steeply
      to zero for $\lambda=0.823$.
      The system is initially in the fine-tuned coherent state with $Q(0) = 0.9049$,
      $P(0) = -1.6382$, $q(0) = -0.0204$, $p(0) = 0.171581$.
      }
    \label{fig:specialcs}
  \end{center}
\end{figure}
In the scenario considered above the system is prepared in the ground
state of the time-independent Dicke Hamiltonian, then suddenly the rotation
is switched on and drives the system out of equilibrium.
In that case we have found a reduced photon number but not a complete
darkening.
We address now the question whether it is possible to fine-tune the
initial conditions in such a way that the dynamically generated
darkness is complete, i.e. $\bar{n}_{\mathrm{ph}}=0$.
To achieve this we simply start at the origin of the coordinate system
and also assume that one of the initial momenta vanishes while the
other is infinitesimally small.
We then follow the evolution for $\lambda=0.823$, the coupling where
the previous protocol has provided a minimum in $\bar{n}_{\mathrm{ph}}$.
The system remains for a long time close to the initial point in phase
space until it moves away and acquires rather large values of both
spatial coordinates and momenta.
Such a point can then be used as initial condition for other values of
$\lambda$.
Fig.~\ref{fig:specialcs} shows both the mean photon number
$\bar{n}_{\mathrm{ph}}$ and the atomic inversion
$\bar{n}_{\mathrm{at}}$ as a function of $\lambda$ for the fine-tuned
starting point
$Q(0) = 0.9049$, $P(0) = -1.6382$, $q(0) = -0.0204$, $p(0) = 0.171581$.
As expected both $\bar{n}_{\mathrm{ph}}$ and $\bar{n}_{\mathrm{at}}$
approach zero for $\lambda=0.823$, or, stated otherwise, there is a
complete slowing down at the origin, as in the case of the Mexican
hat.
Other minima of $\bar{n}_{\mathrm{ph}}$ occur at $\lambda=0.693$ and
$\lambda=0.507$, but $\bar{n}_{\mathrm{at}}$ shows the opposite
behavior at the latter point, it exhibits a maximum instead of a
minimum.
For strong couplings both $\bar{n}_{\mathrm{ph}}$ and
$\bar{n}_{\mathrm{at}}$ approach the stationary values, as predicted
by the solutions of the linearized mean-field equations.

\section{Interplay of Geometric and Dynamic Phases}
\label{compgeomdyn}

In the previous section we have interpreted the dynamically generated
darkness in the framework of time-dependent mean-field theory
and attributed the effect to a slowing down of the
classical motion.
For a system with a small number of two-level atoms mean-field theory
is no longer valid and one has to solve the time-dependent
Schr\"odinger equation.
We believe that in this case the dynamically generated darkness can be
traced back to an interplay between geometric and dynamic phases
appearing away from equilibrium~\cite{TPG}.
To support this claim we represent the time-dependent Schr\"odinger
equation
\be
  i \partial_{t} \ket{\psi_{\mathrm{RD}}(t)} = \op{H}_{\mathrm{RD}}(t) \ket{\psi_{\mathrm{RD}}(t)}
\ee
in the instantaneous basis $\ket{\varphi_{l}(t)}$ of
$\op{H}_{\mathrm{RD}}(t)$, defined by 
\be
  \op{H}_{\mathrm{RD}}(t) \ket{\varphi_{l}(t)} = \epsilon_{l}(t) \ket{\varphi_{l}(t)},
\ee
with $l$ enumerating all the eigenstates of $\op{H}_{\mathrm{RD}}(t)$
at an instant $t$,
e.g. $\ket{\varphi_{1}(t)}$ is the ground state of $\op{H}_{\mathrm{RD}}(t)$ at time
$t$.
The time evolution of the instantaneous eigenstates of
$\op{H}_{\mathrm{RD}}(t)$ is simply given by a rotation of the eigenstates of
$\op{H}_{\mathrm{D}}$, $\ket{\varphi_{l}(t)} = \op{R}_{z}^{\dagger}(t) \ket{\varphi_{l}(0)}$,
with time-independent eigenvalues $\epsilon_{l}(t) = \epsilon_{l}(0)$ 
(those of $\op{H}_{\mathrm{D}}$), because
the rotationally driven Dicke Hamiltonian
$\op{H}_{\mathrm{RD}}(t)$ is obtained by applying the unitary transformation~(\ref{rot_op})
to the time-independent Dicke Hamiltonian $\op{H}_{\mathrm{D}}$.
Inserting the expansion 
$\ket{\psi_{\mathrm{RD}}(t)} = \sum_{l} c_{l}(t) \ket{\varphi_{l}(t)}$ 
into the Schr\"odinger equation yields a
differential equation for the coefficients $c_{l}(t)$
\be
  \frac{d c_{l}}{dt}
  =
  i \lp 
      - \epsilon_{l}
      + A_{ll}
    \rp 
  c_{l}
  +
  i \sum_{k, k \neq l} A_{lk} c_{k},
  \label{eq:dandt}
\ee
where we introduced the notation $A_{lk}(t):= i \bra{\varphi_{l}(t)} \frac{d}{dt} \ket{\varphi_{k}(t)}$.
The integral of the so-called Berry connection
$A_{ll}(t) = i \bra{\varphi_{l}(t)} \frac{d}{dt} \ket{\varphi_{l}(t)}$
~\cite{Xiao2010} is
the geometric phase of the $l$-th instantaneous eigenstate, the
Berry phase~\cite{Berry1984}
\be
  \gamma_{l}(t)
  = 
  \int_{0}^{t}d\tau \, i \, \bra{\varphi_{l}(\tau)} \frac{d}{d\tau} \ket{\varphi_{l}(\tau)}.
  \label{eq:adiberryphase}
\ee
Eq.~(\ref{eq:dandt}) shows that for an adiabatic evolution (where
transitions to different levels $k \neq l$ can be neglected) each instantaneous
eigenstate $\ket{\varphi_{l}(t)}$ acquires not only the well known
dynamic phase $\mathcal{E}_{l}(t) = \int_{0}^{t}d\tau
\epsilon_{l}$ but also a geometric phase $\gamma_{l}(t) =
\int_{0}^{t}d\tau A_{ll}(\tau)$.
Applying the gauge transformation
\be
  c_{l}(t) = \chi_{l}(t) \exp\lp i \int_{0}^{t}d\tau (-\epsilon_{l}+A_{ll}(\tau)) \rp
\ee
to Eq.~(\ref{eq:dandt}) we obtain the equation
\be
  \frac{d \chi_{l}}{dt}
  =
  i \sum_{k, k \neq l} A_{lk}(\phi)
  \exp\lp i \mathcal{E}_{lk}(t) - i \gamma_{lk}(t) \rp
  \chi_{k}(t),
  \label{eq:compgeomdyn}
\ee
which emphasizes the interplay between the dynamic phase
$\mathcal{E}_{lk}(t) = \mathcal{E}_{l}(t)-\mathcal{E}_{k}(t)$ and
the geometric phase $\gamma_{lk}(\phi) =  \gamma_{l}(t)-\gamma_{k}(t)$.
For our model with a constant rotation speed, $\phi(t)=\delta_\phi t$,
the characteristic time scale of the drive is
$T_\phi=2\pi/\delta_\phi$.
After a single rotation the dynamic phase is given by
$\mathcal{E}_{lk}(t) =2\pi(\epsilon_l-\epsilon_k)/\delta_\phi$ and is large for 
$\epsilon_l\neq\epsilon_k$ if $\delta_\phi$ is small.
Thus for adiabatic driving ($\delta_{\phi} \to 0$) the level
transitions in the dynamics are dictated by the dynamic phase.
However, in the case of non-adiabatic driving the geometric phase can
affect the transitions and lead to non-trivial phenomena.

As in the previous section we study the evolution after the
rotation is switched on suddenly at $t=0$, but now we assume the system 
to be prepared in the exact ground state of the time-independent Dicke Hamiltonian
$\op{H}_{\mathrm{D}}$. 
Moreover we now calculate the time evolution of the mean photon number 
using Eq.~(\ref{eq:compgeomdyn}).
The coefficients $\chi_k$ are calculated numerically 
for the initial condition $\chi_{l}(0)=\delta_{l,1}$
and used for evaluating quantities like the photon number
\be
 n_{\mathrm{ph}}(t)
  =
  \frac{1}{j} 
  \sum_{k,l} \chi_{k}^{\ast}(t) \chi_{l}^{\phantom{\ast}}(t) e^{i \mathcal{E}_{kl}(t) - i \gamma_{kl}(t)}
  \bra{\varphi_{k}(0)} \op{a}^{\dagger}\op{a} \ket{\varphi_{l}(0)}.
  \label{eq:nphadi}
\ee
We also get
\be
  A_{lk}(\phi) = -\delta_\phi \bra{\phi_{l}(0)} \op{J}_{z} \ket{\phi_{k}(0)}
\ee
and therefore a geometric phase
\be
  \gamma_{l}(t) 
  =
  \int_{0}^{t} d\tau A_{ll}(\tau)
  =
  - \bra{\varphi_{l}(0)} \op{J}_{z} \ket{\varphi_{l}(0)} \, \phi(t)\, .
  \label{eq:berryphasel}
\ee

\begin{figure}[h!]
  \begin{center}
    \includegraphics[scale=0.45]{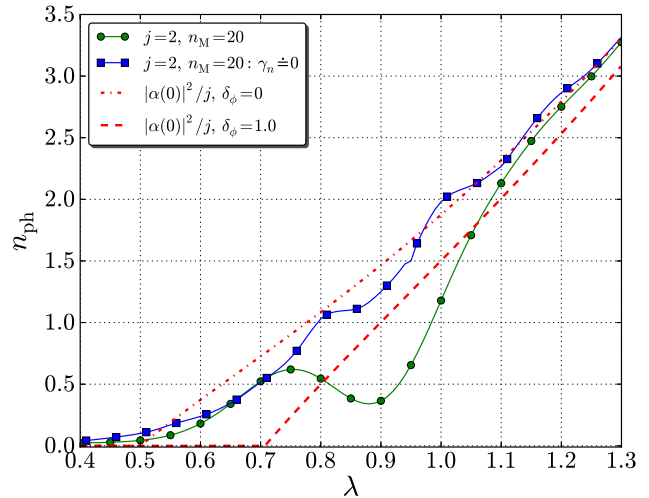}
    \caption{
      Mean photon number after a single period of revolution
      $T_{\phi}$ as a function of the coupling and for
      $\omega=\omega_{0}=\delta_{\phi}=1$.
      The full lines have been obtained by numerically solving Eq.~(\ref{eq:compgeomdyn})
      and using Eq.~(\ref{eq:nphadi}) for $j=2$ with $n_{\mathrm{M}}=20$.
      The blue line (squares) was calculated by setting all the
      geometric phases to zero, $\gamma_{l} = 0, \, \forall \,
      l=1,2,\ldots,(2j+1)(n_{\mathrm{M}}+1)$. 
      Dashed lines represent mean-field values for
      $\op{H}_{\mathrm{D}}$ and $\op{H}_{\mathrm{ROT}}$, respectively.
      }
    \label{fig:adaonetornobp}
  \end{center}
\end{figure}

\begin{figure}[h!]
  \begin{center}
    \includegraphics[scale=0.45]{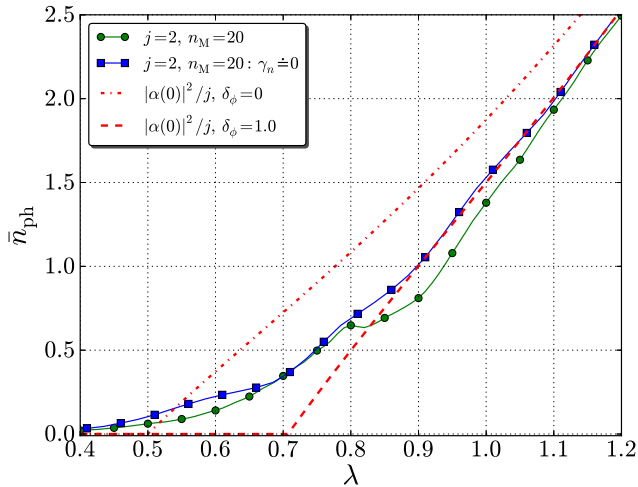}
    \caption{
      Coupling dependence of the mean photon number averaged over a
      time $T=10 \, T_{\phi}$ and for
      $\omega=\omega_{0}=\delta_{\phi}=1$.
      The full lines have been obtained by numerically solving
      Eq.~(\ref{eq:compgeomdyn}) and time-averaging
      Eq.~(\ref{eq:nphadi}) for $j=2$ and $n_{\mathrm{M}}=20$.
      The blue line (squares) was calculated by
      setting all the geometric phases to zero.
      }
    \label{fig:adatavenobp}
  \end{center}
\end{figure}

We have studied the influence of the geometric phases $\gamma_{l}$ on
the dynamics by calculating the mean photon number using Eq.~(\ref{eq:compgeomdyn})
and Eq.~(\ref{eq:nphadi}), once with the values of $\gamma_{l}$ given
in Eq.~(\ref{eq:berryphasel}), once for $\gamma_{l}=0$ for all levels,
$l=1,2,\ldots,(2j+1)(n_{\mathrm{M}}+1)$.
The results for $n_{\mathrm{ph}}(T_{\phi})$, shown in
Fig.~\ref{fig:adaonetornobp}, clearly indicate that the minimum of the
full solution (green line with dots) disappears if the geometric
phases are artificially set to zero (blue line with squares).
The same effect is seen in the time-averaged photon number,
illustrated in Fig.~\ref{fig:adatavenobp}, where the darkening is less
pronounced.
Here we noticed also a difference between the results obtained with
and without geometric phases, at small values of $\lambda$.
We attribute this disparity to finite size effects, which are expected
to be most pronounced for weak coupling.
Unfortunately, calculations based on Eq.~(\ref{eq:compgeomdyn}) are hardly practicable
for $j>2$. 
Therefore we were not able to study size dependence using
this method.

The results of Fig.~\ref{fig:adaonetornobp} and~\ref{fig:adatavenobp}
show that the geometric phases play an important role in the region
slightly above the critical point $\lambda_{c}$.
In fact, both for $\lambda_{c}^{0}<\lambda<\lambda_{c}$
and $\lambda \gg \lambda_{c}$ their effect on the photon density seems
to be at most marginal.
The fact that the minimum disappears if the geometric phases are
quenched suggests that these phases are instrumental in the
dynamically generated darkness.

\section{Conclusion}
\label{sec:conclusion}

We have studied the dynamics of a rotationally driven version of the
Dicke model, where the collective spin is rotated with a constant
velocity about a fixed axis.
The main aim was to calculate the effect of the imposed drive on
important quantities such as the mean photon number and the atomic
inversion.
We have used both a numerical technique for the exact quantum dynamics
of a finite number of two-level atoms and time-dependent mean-field
theory, which is expected to be valid in the thermodynamic limit.
The results obtained with these two methods are consistent with each
other, although the relatively small size treated numerically does not
allow us to draw a definitive conclusion about the limiting behavior
of the exact quantum dynamics for arbitrary large system sizes.

Most calculations have been carried out within a co-rotating frame,
where the dynamics is governed by a time-independent Dicke model with
a renormalized level splitting.
In this frame there exists a ground state which undergoes a transition
as a function of the coupling strength from a normal phase with
vanishing photon number to a superradiant phase with an increasing
number of photons (as well as a growing atomic inversion).
The critical coupling strength is larger than for the undriven Dicke
model and increases with the driving velocity.
In the un-rotated frame the expectation value of the collective spin
precesses around the static field above the critical point, with the
frequency of the drive.

We have studied in great detail the quench dynamics of the driven
Dicke model where the system is prepared in the mean-field ground
state of the undriven Hamiltonian and then suddenly experiences the
rotation.
We found a remarkable dynamically generated darkness, a reduction of
the photon number due to the rotational driving, both in the numerical
calculations for the quantum evolution and in the classical dynamics
of the mean-field theory.
We have interpreted this phenomenon in two different ways.
On the one hand, we have attributed it to a nonlinear slowing down of
the classical dynamical system representing time-dependent mean-field
theory.
On the other hand, for a small number two-level atoms, where the
evolution has to be treated quantum mechanically, we identified the
geometric phase of the instantaneous eigenstates as the main actor
producing the darkening.
It would be very interesting to explore possible connections between
the two interpretations.
Thus one may wonder about the fate of the geometric phase in the
mean-field limit.
For a very simple quantum-mechanical system the geometric phase has
been related to the shift of angle variables in the classical
limit~\cite{Berry1985}, known as Hannay angle~\cite{Hannay}.
This angle has been introduced in the context of slowly
(adiabatically) cycled integrable systems and therefore cannot be
transferred immediately to our model.
Future studies may find out whether some classical analogon for the
geometric phases of quantum dynamics (Berry phases) are responsible
for the dynamically generated darkness reported here.

\section{Acknowledgments}

This work was supported by the Swiss National Science
Foundation. V.G. is grateful to KITP for hospitality.


\begin{appendix}

\section{Solution of the initial value problem for the linearized dynamical system}
\label{A}

We consider the classical dynamical system of
Section~\ref{subsec_linear} and assume the initial deviations from
equilibrium - $Q(0)$ and $q(0)$ for the normal phase, $X(0)$ and
$x(0)$ for the superradiant phase - 
to be small, together with $P(0)=p(0)=0$. 
This allows us to use the linearized equations of motion.

\subsection{Normal phase: ${\mathbf\lambda<\lambda_c}$}

For $\lambda<\lambda_c$ the linearized equations of motion are given
by Eq.~(\ref{eqsmot1}).
In view of the initial conditions $P(0)=p(0)=0$ the general solution is
\beq
Q(t)&=&A_+ \cos (\varepsilon_+ t)+A_- \cos (\varepsilon_- t)\, ,\nonumber\\
q(t)&=&a_+ \cos (\varepsilon_+ t)+a_- \cos (\varepsilon_- t)\, ,
\label{ansatz1}
\eeq
with eigenmode frequencies $\varepsilon_\pm$ given by
Eq.~(\ref{eigenfr1}).
Eqs.~(\ref{eqsmot1}) imply the following relation between coefficients
\be
 a_{\pm}
 =
 f_{\pm} A_{\pm}, \quad \mbox{with} \quad 
 f_{\pm} = \frac{\varepsilon_\pm^2-\Omega^2}{2\lambda\Omega}\, .
\ee
The initial conditions yield the amplitudes
\be
 A_{+}
 =
 \frac{q(0)-Q(0)f_-}{f_+-f_-}\, ,\qquad A_-=\frac{Q(0)f_+-q(0)}{f_+-f_-}\, .
\label{coeff1}
\ee

\subsection{Superradiant phase: ${\mathbf \lambda>\lambda_c}$}

For $\lambda>\lambda_c$ the linearized equations of
motion~(\ref{eqsmot2}) describe the deviations $X(t)=Q(t)-Q_0$,
$x(t)=q(t)-q_0$ from the stationary state $Q_0,\, q_0$.
The general solution for $P(0)=p(0)=0$ is
\beq
X(t)&=&A_+ \cos (\varepsilon_+ t)+A_- \cos (\varepsilon_- t)\, ,\nonumber\\
x(t)&=&a_+ \cos (\varepsilon_+ t)+a_- \cos (\varepsilon_- t)\, 
\label{ansatz2}
\eeq
with eigenmode frequencies given by Eq.~(\ref{eigenfr2}).
Using Eqs.~(\ref{eqsmot2}) we find the following relation between
coefficients
\be
 a_{\pm}
 =
 g_{\pm} A_{\pm}, \qquad 
 g_{\pm}
 =
 \frac{2\sqrt{2}\, \omega\lambda_c^2}
{(\varepsilon_\pm^2 - \omega^2)\sqrt{\lambda^2+\lambda_c^2}}\, ,
\ee
while the initial conditions imply 
\be
 A_{+}
 =
 \frac{x(0)-X(0)g_-}{g_+-g_-}\, ,\qquad 
 A_{-}
 =
 \frac{X(0)g_+-x(0)}{g_+-g_-}\, .
\label{coeff2}
\ee
For $\lambda\gg\lambda_c$ the coefficient $g_+$ is very small and 
$x(t)\approx a_-\cos(\varepsilon_-t)$, where $\varepsilon_-\approx\omega$.

\subsection{Mean photon number}

It is now straightforward to calculate some important quantities such
as the mean photon number.
For simplicity we only consider values averaged over long times and calculate 
\be
\overline{n}_{\mbox{\scriptsize ph}}
=
\lim_{T\rightarrow\infty} \frac{1}{2T}\int_0^T\, dt\, [q^2(t)+p^2(t)]\, .
\ee
We choose the initial conditions of Section~\ref{darkness},
i.e. $Q(0)$ and $q(0)$ correspond to the stationary state of the
time-independent Dicke model.
We limit ourselves to the nontrivial case  $\lambda>\lambda_c^0$, where
\beq
Q(0)&=&\pm \sqrt{2}\left(1-\frac{[\lambda_c^0]^2}{\lambda^2}\right)^{\frac{1}{2}}\, ,\nonumber\\
q(0)&=&\mp\frac{2\lambda}{\omega}\left(1-\frac{[\lambda_c^0]^4}{\lambda^4}\right)^\frac{1}{2}\, .
\label{initial}
\eeq 

For $\lambda_c^0<\lambda<\lambda_c$ the system evolves in the normal
phase of the time-dependent Dicke model, where the stationary values
of the coordinates vanish, $Q_0= q_0=0$. 
We obtain
\be
\overline{n}_{\mbox{\scriptsize ph}}
=
\frac{1}{4\omega^2}[(\varepsilon_+^2+\omega^2)a_+^2+(\varepsilon_-^2+\omega^2)a_-^2] \, ,
\ee
where the eigenvalues $\varepsilon_\pm$ and coefficients $f_\pm$ are
given by Eqs.~(\ref{eigenfr1}) and (\ref{coeff1}), respectively.

For $\lambda>\lambda_c$, i.e. in the superradiant phase of the
time-dependent Dicke model, the photon number of the ground state is
non-zero and the stationary coordinates $Q_0, q_0$ are given by
Eq.~(\ref{stationary}).
The time-dependent coordinates are $q(t)=q_0+x(t)$ and
$Q(t)=Q_0+X(t)$.
Therefore the photon number averaged over a long time $T$ is 
\be
\overline{n}_{\mbox{\scriptsize ph}}
=
\frac{1}{2}q_0^2+\frac{1}{4\omega^2}[(\varepsilon_+^2+\omega^2)a_+^2+(\varepsilon_-^2+\omega^2)a_-^2]\, ,
\ee
where $\varepsilon_\pm$ and $a_\pm$ are defined by
Eqs.~(\ref{eigenfr2}) and (\ref{coeff2}), respectively, 
and $x(0)=q(0)-q_0$, $X(0)=Q(0)-Q_0$.
\begin{figure}[h!]
  \begin{center}
    \includegraphics[scale=0.6]{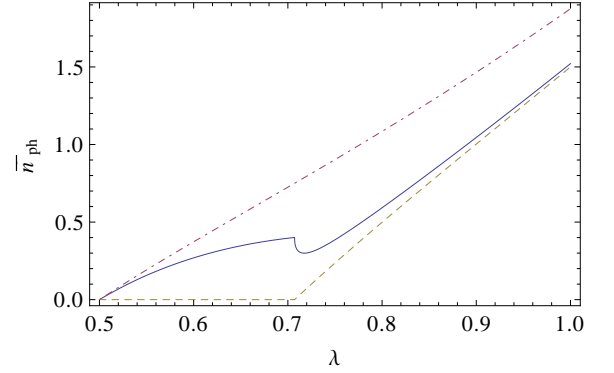}
    \caption{Mean photon number in the linear limit for initial
      conditions corresponding to the ground state of the
      time-independent Dicke Hamiltonian and for a subsequent
      evolution governed by the time-dependent Dicke Hamiltonian with
      parameters $\omega=\omega_{0}=1, \delta_\phi = 1$.
      }
    \label{fig:nph_lin}
  \end{center}
\end{figure}
The result shown in Fig.~\ref{fig:nph_lin} is in qualitative agreement
with that obtained with the full Hamiltonian
(Fig.~\ref{fig:adalamtascs}), although the linear approximation cannot
be trusted for $\lambda$ of the order of $\lambda_c$, where the
potential is strongly nonlinear.

\section{Slowing down in the Mexican hat}
\label{B}

\begin{figure}[h!]
  \begin{center}
    \includegraphics[scale=0.6]{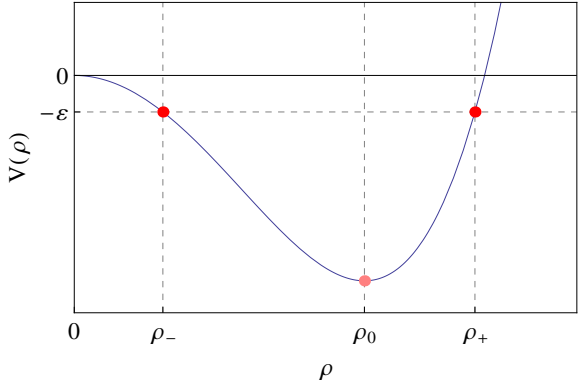}
    \caption{
      Potential as a function of the radius $\rho$.
      For a negative energy $-\varepsilon$, the motion is limited by
      turning points $\rho_\pm$ and does not reach the origin. 
      }
    \label{fig:potential_mex}
  \end{center}
\end{figure}

\begin{figure}[h!]
  \begin{center}
    \includegraphics[scale=0.6]{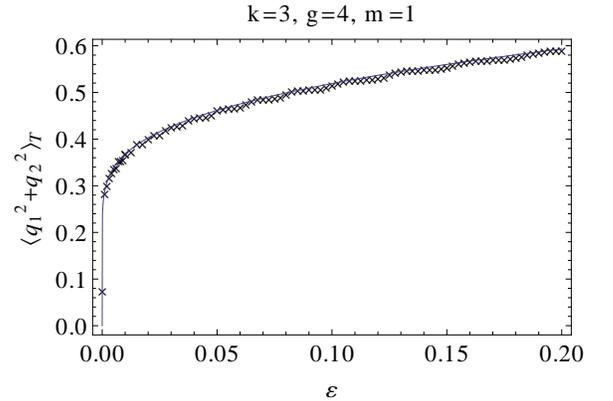}
    \caption{Time average of $q_{1}^{2}(t)+q_{2}^{2}(t)$ at negative
      energies $-\varepsilon$ for the isotropic ``Mexican hat''
      potential~(\ref{eq:pot_mex}) with parameters $m=1$, $k=3$,
      $g=4$. 
      The full line is the analytical solution, crosses stand for the
      numerical integration of the equations of motion.
      }
    \label{fig:darkness_mex}
  \end{center}
\end{figure}

To illustrate the dynamically generated darkness detected for a finite
rotation velocity above the critical point of the Dicke model
(Section~\ref{darkness}) we consider a classical system of two degrees
of freedom described by the Lagrangian
\be
 L
 =
 \frac{m}{2}(\dot{q}_1^2+\dot{q}_2^2)+\frac{k}{2}(q_1^2+q_2^2)-\frac{g}{4}(q_1^2+q_2^2)^2\, ,
\label{lagr_mexhat}
\ee
where $k>0$, $g>0$.
In polar coordinates, $q_1=\rho\cos\varphi$, $q_2=\rho\sin\varphi$,
Eq.~(\ref{lagr_mexhat}) reads
\be
 L
 =
 \frac{m}{2}(\dot{\rho}^2+\rho^2\dot{\varphi}^2)-V(\rho)\, ,
\ee
where  
\be
 V(\rho)
 =
 -\frac{k}{2}\rho^2+\frac{g}{4}\rho^4
\label{eq:pot_mex}
\ee
has a minimum $V_{\mbox{\scriptsize min}}=-k^2/(4g)$ at
$\rho_0=\sqrt{k/g}$.
For initial conditions $\dot{q}_1(0)=\dot{q}_2(0)=0$ a particle moving
in this ``Mexican hat'' executes radial motions, around $\rho_0$ if
the energy is negative and centered at $\rho=0$ if the energy is positive.
In either case the particle spends a lot of time at small values of
$\rho$ if the energy is small.
To see this quantitatively we consider the special case of a negative
energy (for $\dot{\varphi}=0$),
\be
\frac{m}{2}\dot{\rho}^2-\frac{k}{2}\rho^2+\frac{g}{4}\rho^4=-\varepsilon
\label{neg_energy}
\ee
with $\varepsilon>0$, where the particle moves between turning points 
\be
\rho_\pm^2=\rho_0^2\pm\sqrt{\rho_0^4-\frac{4\varepsilon}{g}}\, ,
\ee
as illustrated in Fig.~\ref{fig:potential_mex}.
Eq.~(\ref{neg_energy}) then implies
\be
dt=\pm\sqrt{\frac{2m}{g}}\frac{d\rho}{\sqrt{(\rho^2-\rho_-^2)(\rho_+^2-\rho^2)}}\, .
\ee
The time $T$ for moving from $\rho_-$ to $\rho_+$ (or the other way) is
\begin{align}
T
&=
\sqrt{\frac{2m}{g}}\int_{\rho_-}^{\rho_+}\frac{d\rho}{\sqrt{(\rho^2-\rho_-^2)(\rho_+^2-\rho^2)}} \nonumber \\
&=
\sqrt{\frac{2m}{g}}\frac{1}{\rho_+}K\left(\frac{\sqrt{\rho_+^2-\rho_-^2}}{\rho_+}\right)\, ,
\end{align}
and similarly we obtain
\begin{align}
\int_0^T dt\, \rho^2(t)
&=
\sqrt{\frac{2m}{g}}\int_{\rho_-}^{\rho_+}\frac{\rho^2d\rho}{\sqrt{(\rho^2-\rho_-^2)(\rho_+^2-\rho^2)}} \nonumber \\
&=\sqrt{\frac{2m}{g}}\rho_+E\left(\frac{\sqrt{\rho_+^2-\rho_-^2}}{\rho_+}\right)\, ,
\end{align}
where $E(k)$ and $K(k)$ are complete elliptic integrals. Therefore the time-average of
$\rho^2$ is given by
\be
\langle\rho^2\rangle_T:=\frac{1}{T}\int_0^T dt\, \rho^2(t)
=\frac{E\left(\frac{\sqrt{\rho_+^2-\rho_-^2}}{\rho_+}\right)}
{K\left(\frac{\sqrt{\rho_+^2-\rho_-^2}}{\rho_+}\right)}\, \rho_+^2\, .
\label{rho_mexhat}
\ee
For $\varepsilon \rightarrow 0$ the turning point $\rho_-$ moves to 0
and the elliptic integrals have the limiting behavior 
\be
E\left(\frac{\sqrt{\rho_+^2-\rho_-^2}}{\rho_+}\right)\sim 1, \quad 
K\left(\frac{\sqrt{\rho_+^2-\rho_-^2}}{\rho_+}\right)\sim\log\left(\frac{4\rho_+}{\rho_-}\right)\, .
\ee
It follows that the average value of $\rho^2(t)$ tends to zero if
$\varepsilon$ tends to zero, i.e. if the initial potential energy
coincides with the hump at $\rho=0$.
A similar result is obtained for the average 
$\langle p_\rho^2\rangle_T=m^2\langle \dot{\rho}^2\rangle_T$, which
also tends to zero (logarithmically) for $\varepsilon \rightarrow 0$.
The origin of this behavior is simply that with decreasing energy the
particle slows more and more down by approaching the turning point
$\rho_-$.
At the critical energy (here $\varepsilon=0$), where the motion
changes from a simple oscillation about a minimum (for
$\varepsilon<0$) to a motion over the potential hump (for $\varepsilon
>0$) the time span for a single period diverges and both
$\langle\rho^2\rangle_{T}$ and $\langle p_\rho^2\rangle_{T}$ tend to zero.
Fig.~\ref{fig:darkness_mex} shows both the analytical
expression~(\ref{rho_mexhat}) and numerical results for
$\langle\rho^2\rangle_{T}$.
The agreement is excellent and serves as a test for the algorithm used
in the numerical calculations.

\end{appendix}


\end{document}